\documentstyle[amssymb,aps,twocolumn]{revtex}

\begin{document}
\title{Capture Velocity for a Magneto-Optical Trap in a Broad Range of Light
Intensity}
\author{S. R. Muniz, K. M. F. Magalh\~{a}es, Ph. W. Courteille, M. A. Perez, L. G.
Marcassa and V. S. Bagnato}
\address{Instituto de F\'{i}sica de S\~{a}o Carlos, Universidade de S\~{a}o Paulo,\\
Caixa Postal 369, 13560-970, S\~{a}o Carlos-SP Brazil.}
\maketitle

\begin{abstract}
{\em In a recent paper, we have used the dark-spot Zeeman tuned slowing
technique [Phys. Rev. A 62, 013404-1, (2000)] to measure the capture
velocity as a function of laser intensity for a sodium magneto optical trap.
Due to technical limitation we explored only the low light intensity regime,
from }$0\ ${\em to }$27\;mW/cm^{2}${\em . Now we complement that work
measuring the capture velocity in a broader range of light intensities (from 
}$0\ ${\em to }$400\;mW/cm^{2}${\em ). New features, observed in this range,
are important to understant the escape velocity behavior, which has been
intensively used in the interpretation of cold collisions. In particular, we
show in this brief report that the capture velocity has a maximum as
function of the trap laser intensity, which would imply a minimum in the
trap loss rates.}

PACS number(s): 32.80.Pj, 33.80.Ps, 34.50.Rk, 34.80.Qb
\end{abstract}

\bigskip

\section{Introduction}

In the prediction and the interpretation of trap loss rates, an important
parameter used is the escape velocity ($\upsilon _{e}$). Which is defined as
the minimum velocity an atom has to achieve in order to overcome the
radiative forces to escape from the trap. This velocity is normally obtained
through numerical calculations where a single atom is considered within the
trap light forces. Such simulations has been satisfactory in many cases, but
real measurements are important to reveal facts not included in the
simulations. The direct measurements of $\upsilon _{e}$ would be ideal, but
so far nobody yet has devised a direct scheme to perform such experiments.
So what has been actually measured is the so called capture velocity ($%
\upsilon _{c}$), defined as the maximum velocity of an atom traversing the
trap volume that can still be captured into the trap. The value of $\upsilon
_{c}$ is related to the value of $\upsilon _{e}$, and $\upsilon _{c}$ can be
considered as an upper bound limit for $\upsilon _{e}$.

In a recent publication \cite{b1} we have used a magneto optical trap loaded
from a dark-spot Zeeman tuned slower \cite{b2}\ to measure the capture
velocity in a range of trapping light intensities from $0$ to $27$ $%
mW/cm^{2} $. This range of intensity covers only one part of the interest
when dealing with measurements of trap loss. Recently collision experiments 
\cite{b3} have covered a much broader range of intensities and the
prediction for trap loss in this range demands the knowledge of $\upsilon
_{e}$. High intensity is however a difficult regime to simulate because
effects such as fluctuations, polarization imperfection, the multilevel
nature of atoms, etc. are hard to be included in the calculation.
Measurements of $\upsilon _{c}$ in this range, may give us a good idea about 
$\upsilon _{e}$ for the interpretation of collision experiments.

In what follows, we provide a brief description of our experimental
technique followed by our most recent measurement of $\upsilon _{c}$ and a
plausible explanation for the observed behavior. This report should be
considered as a complement to our previous work \cite{b1}.

\section{\protect\bigskip Experimental setup}

To load our sodium MOT we use a cold atomic beam from a Zeeman slower where
a dark spot is placed in the center of the slowing laser beam producing a
shadow on the position of the trap. This results in a large capture of atoms
in the MOT, due to the the minimum disturbing effect of the slowing laser on
the atoms in the MOT region. This technique has been fully described in ref. 
\cite{b2}. Scanning the slowing laser, different velocity classes appear in
the outgoing beam of atoms. The distribution is narrow and the dependence on
the slowing laser frequency and its width has been studied before for a wide
range of laser frequencies\cite{b1}. The peak velocity depends on the
slowing laser frequency and basically obeys the relation $k\upsilon _{out}$
= - $\Delta $\cite{b3}, where $\Delta $ is the slowing laser detuning and $k$
is the wave vector. This outgoing velocity peak has also been investigated
as a function of detuning of the slowing laser \cite{b1}.

The output velocity distribution from the slowing process is close to a
Gaussian distribution $g(\upsilon )=Ae^{-(\upsilon +\Delta /k)^{2}/\sigma
^{2}}$, where A is a normalizing constant and $\sigma =\Delta \upsilon /2\,\
\,$is obtained from the previous investigations \cite{b1} of the atomic
distribution coming out from the slowing process.

The procedure to measure the capture velocity is as follows: the slowing
laser is scanned ( $\thicksim $5MHz/s) as the number of \ trapped atoms is
recorded. If the laser frequency is far to the red, the atoms cross the trap
volume too quickly and will not be captured. When the frequency is such that
small velocities are present, atoms start to accumulate in the trap until
the whole distribution is captured. The number of captured atoms for each
laser detuning is well represented by the integral \cite{b1}

\bigskip 
\begin{equation}
N_{c}=\int_{0}^{v_{c}}g(\upsilon )d\upsilon
\end{equation}

\begin{equation}
N_{c}(\Delta )=A\int_{0}^{\upsilon _{c}}e^{-(\upsilon +\Delta /k)^{2}/\sigma
^{2}}d\upsilon  \label{eq2}
\end{equation}

This integral can be expressed as a combination of error functions, which
provides an analytical expression for $N_{c}$($\Delta $). This function is
however very dependent on $\upsilon _{c}$. In fig. 1 we show $N_{c}$($\Delta 
$) normalized for several considered capture velocities. As the capture
velocity increases, the curves for $N_{c}$($\Delta $) are shifted towards
higher values of negative detunings. To determine $\upsilon _{c}$, we have
fitted the rising part of the measured spectrum with the best value of $%
\upsilon _{c}$. The process is repeated for several trap laser intensities
and $\upsilon _{c}$ is obtained as a function of \ trap laser intensity. We
used a trap laser detuning of $\ -10\;MHz\;$and a MOT field gradient of $%
\frac{dB}{dz}\sim 20\;G/cm$. In figure 2 we show our measurement of the
capture velocity for a broad range of intensities. The dot line connecting
the experimental points is only for eye guidance. The uncertainties comes
from \ the statistical fluctuations of the values obtained in the best
fittings, due to the data scattering of different measurements.

The behavior of $\upsilon _{c}$ as a function of trapping laser intensity
for low intensities is as expected: seems to go to zero for zero intensity
and as the intensity increases it reaches a smaller rate of increasing.
This, in fact, agrees well with our previous result \cite{b1}. The new
feature here is that $\upsilon _{c}$ reaches a maximum value for a given
intensity and start to decrease slowly after that point. In our case, the
maximum seems to be around $50\;mW/cm^{2}$. Since this is the total laser
intensity when the six beams are considered, the maximum in $\upsilon _{c}$
is taking place at about $8\;mW/cm^{2}$ per beam (not far from saturation
intensity for each beam). This results seems to confirm the discussions in
ref. \cite{b4}, that the capture process might depend more on the damping
part of the radiation pressure than in the restoring force of the trap. In
this case, the maximum capture would correspond to the situation of optimum
balance between the beams. And any further increase in intensity
oversaturate the transitions and the power broadening thereafter compromises
the atom distinction between the two counterpropagating laser beams. In such
a situation the damping coefficient starts to decrease and the same occurs
to the capture velocity. So far, after the peak capture velocity, we only
observe a decrease of $\upsilon _{c}$ with intensity.

\section{Conclusion}

The main consequence of this observation is that the escape velocity ($%
\upsilon _{e}$) should follow a similar pattern. And such a behavior of the
escape velocity with respect to the trapping laser intensity would mean that
the effective trap depth of a MOT has also a maximum, related with that
position of $\upsilon _{e}$. This maximum in the trap depth would imply the
existence of a minimum for the trap loss rate, which would be consistent
with the recent alternative interpretation for the behavior of the loss rate
coefficient in MOTs as a function of laser intensity \cite{b4}. We should
emphasize, however, that such a dependence is very much related to the
overall condition for the trap operation

\section{Acknowledgments}

The authors wish to thank Dr. John Wiener for his critical reading of the
original manuscript. This work has received support from Fapesp (Programa de
Centros CEPID - Centro de Pesquisa em \'{O}ptica e Fot\^{o}nica CePOF) and
Programa Pronex-CNPq.

\section{\protect\bigskip Figure Captions}

\begin{figure}[tbp]
\caption{Theoretical curves showing the number of trapped atoms as a
function of the slowing laser frequency. As the capture velocity is
increased, the curves shift to higher values of negative detuning.}
\end{figure}

\bigskip

\begin{figure}[tbp]
\caption{ Measured capture velocity as a function of trapping laser
intensity for a detuning $\Delta \sim $ -$10\;Mhz$, half waist of the
Gaussian beam of $4.5mm$ and $\frac{dB}{dz}=\;20\;Gauss/cm$. The dotted line
used to connect the experimental points is just for eye guidance.}
\end{figure}

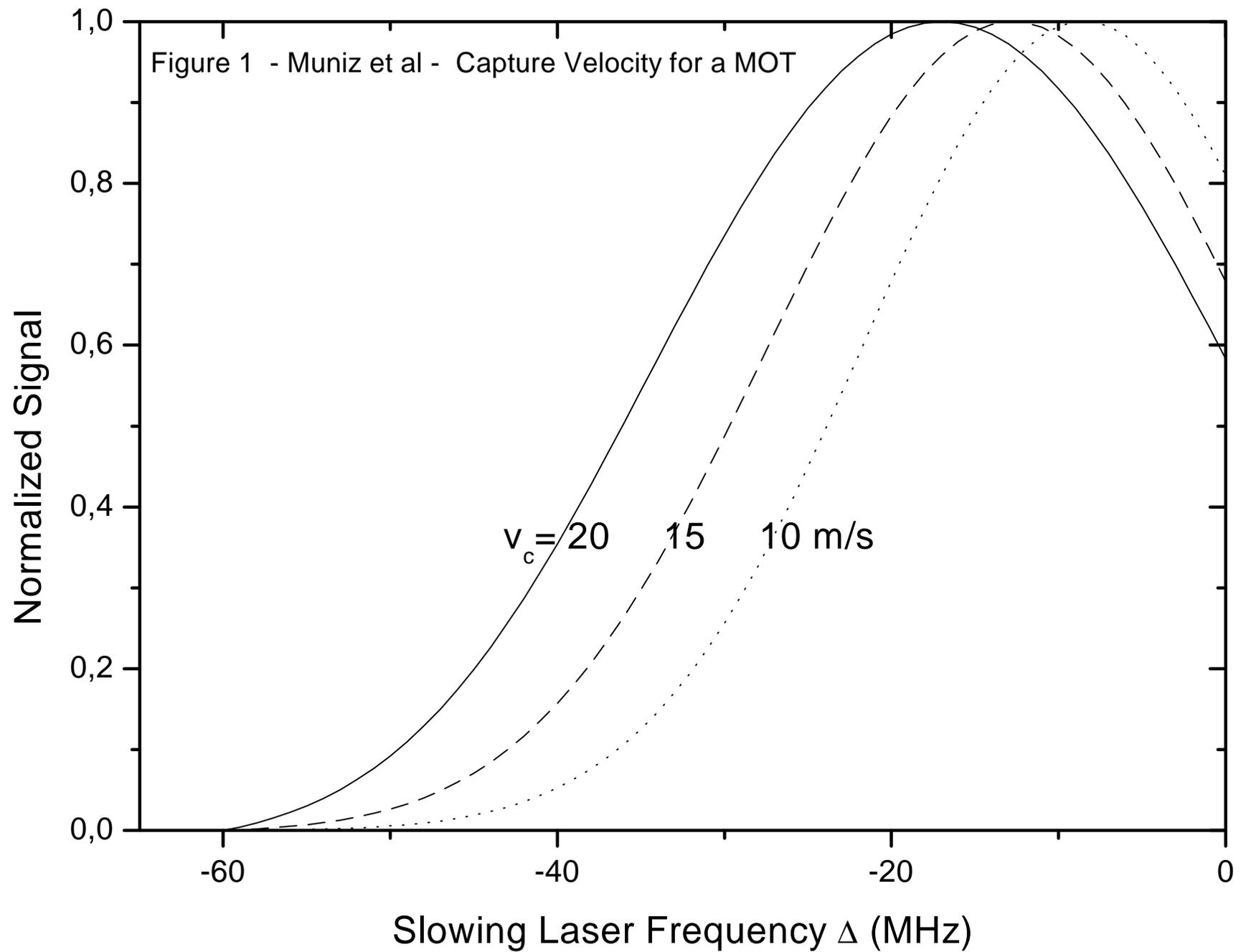
\end{document}